# Extraction of SSVEPs-based Inherent Fuzzy Entropy Using a Wearable Headband EEG in Migraine Patients

Zehong Cao, *Member, IEEE*, Chin-Teng Lin*, *Fellow, IEEE*, Kuan-Lin Lai, Li-Wei Ko, *Member, IEEE*,

Jung-Tai King, Jong-Ling Fuh, Shuu-Jiun Wang*

*Abstract* - Inherent fuzzy entropy is an objective measurement of electroencephalography (EEG) complexity, reflecting the robustness of brain systems. In this study, we present a novel application of multi-scale relative inherent fuzzy entropy using repetitive steady-state visual evoked potentials (SSVEPs) to investigate EEG complexity change between two migraine phases, i.e. inter-ictal (baseline) and pre-ictal (before migraine attacks) phases. We used a wearable headband EEG device with O1, Oz, O2 and Fpz electrodes to collect EEG signals from 80 participants (40 migraine patients and 40 healthy controls [HCs]) under the following two conditions: during resting state and SSVEPs with five 15-Hz photic stimuli. We found a significant enhancement in occipital EEG entropy with increasing stimulus times in both HCs and patients in the inter-ictal phase but a reverse trend in patients in the pre-ictal phase. In the 1$^{st}$ SSVEP, occipital EEG entropy of the HCs was significantly higher than that of patents in the pre-ictal phase (FDR-adjusted $p < 0.05$). Regarding the transitional variance of EEG entropy between the 1$^{st}$ and 5$^{th}$ SSVEPs, patients in the pre-ictal phase exhibited significantly lower values than patients in the inter-ictal phase (FDR-adjusted $p < 0.05$). Furthermore, in the classification model, the AdaBoost ensemble learning showed an accuracy of 81$\pm$6% and AUC of 0.87 for classifying inter-ictal and pre-ictal phases. In contrast, there were no differences in EEG entropy among groups or sessions by using other competing entropy models, including approximate entropy, sample entropy and fuzzy entropy on the same dataset. In conclusion, inherent fuzzy entropy offers novel applications in visual stimulus environments and may have the potential to provide a pre-ictal alert to migraine patients.

*Index Terms* - Migraine, SSVEP, EEG, Inherent Fuzzy Entropy

## I. Introduction

Migraine is a type of neurovascular headache that presents a severe throbbing head pain and is accompanied by nausea, vomiting or extreme sensitivity to light and sound [1]. Episodic migraine is considered a recurrent headache with a cycle that includes inter-ictal, pre-ictal, ictal and post-ictal phases [2]. The ictal phase is the period during which migraine patients suffer from headache; the pre-ictal phase is defined as 72 hours before the ictal phase, which is preceded by the inter-ictal phase.

The activation and sensitization of brain activities, which have powerful potential applications, can be characterized by the visual stimulus environment [3]. In particular, steady-state visual evoked potentials (SSVEPs) are responses to photic stimuli at a multiple of or equal to the frequency of the stimuli. The visual system of the brain is intimately connected with the environment via the eyes, which contain light receptors on the retina. These receptors send electrical signals to the brain, which can be detected by cortical electroencephalography (EEG). These messages are generally sent to the occipital region [4], which causes cortical activation directing to specific frequency stimulations [5]. Migraine patients are more vulnerable to visual stimuli, and previous studies have reported that visual stimuli are effective for examining the dishabituation in the inter-ictal or pre-ictal phases [6-11]. In previous SSVEP studies, it has been shown that the cortical responses would be enhanced in patients with migraine as compared to those of control subjects, in a broad range of stimulation frequency, supporting the concept of central hyper-responsiveness in patients with migraine. Initially, stimulation frequencies at median-to-high range (i.e., 15 - 30 Hz) were demonstrated to have such differentiating property [12, 13]. Lately, it was realized that stimulations at lower frequencies (i.e., 3 - 6 Hz) were also able to elicit such

This work was partially supported by grants from the Australian Research Council under discovery projects [DP180100670 and DP180100656], Army Research Laboratory [W911NF-10-2-0022 and W911NF-10-D-0002/TO 0023], Ministry of Science and Technology of Taiwan [MOST 107-2321-B-010-001 -106-2321-B-010-009 -MOST 104-2745-B-010-003-MOST 103-2321-B-010-017-], Brain Research Center, National Yang-Ming University from The Featured Areas Research Center Program within the framework of the Higher Education Sprout Project by the Ministry of Education (MOE) in Taiwan.

Z. Cao and C.T. Lin are with the Computational Intelligence and Brain Computer Interface Lab, Center for Artificial Intelligence and Faculty of Engineering and Information Technology, University of Technology Sydney, Australia. (*corresponding to Chin-Teng.Lin@uts.edu.au).

L.W. Ko and J.T. King are with Brain Research Center, National Chiao Tung University, Taiwan.

K.L. Lai, J.L. Fuh and S.J. Wang are with Taipei Veterans General Hospital and also with the Brain Research Center and School of Medicine, National Yang-Ming University, Taipei, Taiwan. (*corresponding to sjwang@vghtpe.gov.tw).

differences [14]. Considering overt visual stimulation could evoke migraine attacks, or even provoke seizures [15], we thus decided to use only one stimulation frequency (i.e., 15 Hz), which was well located in the middle of the frequency distribution, to avoid the possible adverse effects.

To efficiently extract features from SSVEP-based models, multivariate linear regression [16] and multi-set/-layer/-way canonical correlation analysis [17-19], which were developed and focused on frequency recognition. In addition to the above measurements, recently developed entropy analysis approaches in the temporal domain have helped us to understand brain dynamics and allowed us to assess how complexity provides information about a wide range of physiological systems [20]. Entropy is generally an objective measure of how the complexity of physiological signals represents the robustness of brain systems [21]. Different entropy analysis approaches, such as approximate entropy (ApEn) [22] and sample entropy (SampEn) [23] have been developed to measure complex signals.

Fuzziness, featured by uncertainty, can avoid a sharp distinction of the boundary of a set [24]. For example, the fuzzy entropy (FuzzEn) algorithm [25, 26] incorporates a fuzzy membership function instead of the Heaviside function to assess the degree of similarity between two vectors' shapes, which can effectively overcome the deficiency in existing subspace filtering techniques [27]. Integrated brain systems are often multi-scaled and can interact with fast or slow processes, depending on the scale of the bio-signal of interest. Thus, we recently developed a multi-scale inherent fuzzy entropy (Inherent FuzzEn) algorithm [28] that has the robustness to operate under noise, nonlinear and non-stationary signals and is capable of operating on EEG signals across a range of temporal (time) scales. Compared to ApEn, SampEn or FuzzEn, the Inherent FuzzEn algorithm shows stable complexity and the smallest root mean square deviation in the resting-state condition. Therefore, we applied the Inherent FuzzEn algorithm to explore resting-state EEG complexity before migraine attacks [29], which showed that the EEG complexity of patients in the pre-ictal phase was significantly higher than that of patients in the inter-ictal phase in the prefrontal area.

To the best of our knowledge, high and low complexity represent healthy (robust) and diseased (vulnerable) brain systems [20, 29]. We hypothesize that a healthy brain system demonstrates strong robustness after repetitive visual stimulation (linked to habituation) [30] but that migraine patients may exhibit less robustness of the brain system. However, existing studies have not investigated the brain complexity in migraine patients as well as healthy controls (HCs) using entropy analysis approaches in repetitive visual stimulus sessions. Thus, we aimed to 1) understand the effects of repetitive SSVEPs in migraine patients and HCs by measuring inherent fuzzy entropy; 2) classify inter-ictal and pre-ictal migraine phases using input features of inherent fuzzy entropy; 3) compare our method to other entropy algorithms (ApEn, SampEn and FuzzEn). Our study represents a novel application of inherent fuzzy entropy in a visual stimulus environment for migraine patients, and the extracted features have the potential to provide pre-ictal alerts to migraine patients.

This paper is organized as follows. First, a relative inherent fuzzy entropy algorithm that minimizes individual differences is presented in Section II. Section III introduces the experimental procedures, including the recruitment of participants, the wearable EEG device, the experimental paradigm, the data processing classification models and the statistical analysis. In Section IV, the application performances of the relative inherent fuzzy entropy algorithm are assessed regarding migraine patients and HCs. Section V discusses the experimental results. Section VI provides the conclusion.

## II. EEG COMPLEXITY: RELATIVE MULTI-SCALE INHERENT FUZZY ENTROPY

EEG complexity is measured by the relative multi-scale inherent fuzzy entropy algorithm [28], which is divided into three parts: A. a de-trending process; B. a multi-scale procedure; and C. fuzzy entropy assessment. In this study, to minimize individual differences, we estimated the differences in multi-scale inherent fuzzy entropy between the baseline (resting) and stimulus sessions (refer to part D), termed the relative multi-scale inherent fuzzy entropy (RE). As shown in Fig. 1, we present an overview of our proposed relative multi-scale inherent fuzzy entropy.

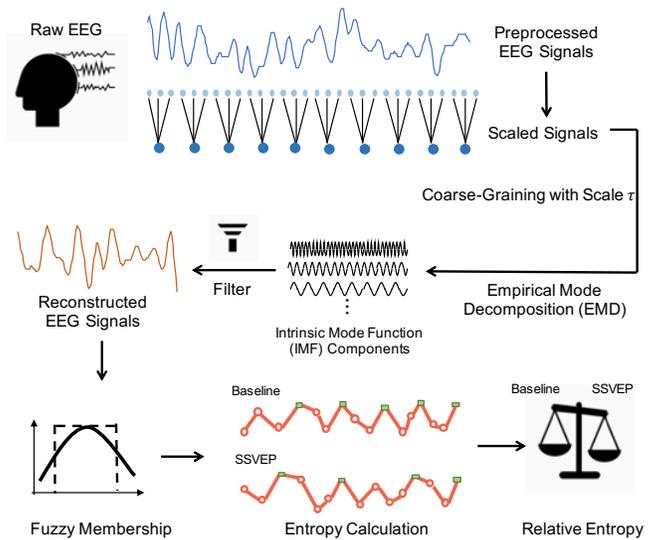

**Figure 1** An overview of relative multi-scale inherent fuzzy entropy

## A. De-trending process to extract inherent modes

We applied empirical mode decomposition to decompose the raw EEG signal $s(t)$ into several intrinsic mode functions and reconstructed the signal $\hat{s}(t)$.

In the initial step, extrema of the signal $s(t)$ are found, corresponding to $E_{minima}$ and $E_{maxima}$. Then, the regions between $E_{minima}$ and $E_{maxima}$ are interpolated, yielding an envelope with $en_{min}(t)$ and $en_{max}(t)$.

First, we compute the mean:

$$M(t) = (en_{min}(t) + en_{max}(t))/2 \quad (1)$$

Second, we extract the candidate of inherent functions:

$$Ca(t) = s(t) - M(t) \quad (2)$$

Third, we confirm $Ca(t)$, belonging to an intrinsic mode function. If $Ca(t)$ satisfies the constraint conditions, $Ca(t)$ is saved, and the residue is computed:

$$res(t) = s(t) - \sum_{i=1}^{t} Ca(t) \quad (3)$$

Next, we solve $t = t + 1$ and treat $res(t + 1)$ as input data. Otherwise, we treat $Ca(t + 1)$ as input data. Iterations are performed on the residual $res(t)$ and continued until the final residue $r$ satisfies the stopping criterion.

Finally, the components of intrinsic mode functions surviving high trends are automatically removed by a trend filter. The signal $\hat{s}(t)$ is reconstructed by the cumulative sum of the remaining intrinsic mode functions:

$$\hat{s}(t) = \sum_{i=n}^{i=m} Ca(t) \quad (4)$$

The parameter $i$ is the order number of the components from the intrinsic mode functions, and parameters $m$ and $n$ are the upper and lower boundaries of the selected components, respectively.

## B. Multi-scale procedure

The initial step is to normalize the EEG signal using the Z-score measurement. The EEG signal $\hat{s}(t)$ subtracts the mean prior to dividing by the standard deviation. The normalized EEG signal is marked as $x(t)$. Afterwards, the multi-scale procedure involves coarse-graining the signals into different time scales.

For a given time series, multiple coarse-grained time series are constructed by averaging the data points within non-overlapping windows of increasing lengths, and the $\tau$ element of the coarse-grained time series $\hat{x}_j^{(\tau)}$ is expressed as:

$$\hat{x}_j^{(\tau)} = \frac{1}{\tau} \sum_{i=(j-1)\tau+1}^{j\tau} x_i \quad (5)$$

where $\tau$ represents the scale factor, and $1 \leq j \leq N/\tau$.

## C. Fuzzy entropy assessment

First, considering the $N$ sample time series $\{\hat{x}(i): 1 \leq i \leq N\}$, given $m$, $n$, and $r$, a vector set sequence $\{X_i^m, i = 1, \dots, N - m + 1\}$ is calculated, and the baseline is removed:

$$X_i^m = \begin{Bmatrix} \hat{x}(i), \hat{x}(i+1) \dots, \\ \hat{x}(i+m-1) \end{Bmatrix} - m^{-1} \sum_{j=0}^{m-1} \hat{x}(i+j) \quad (6)$$

where $1 \leq i \leq N - m + 1$, and $X_i^m$ presents $m$ consecutive $u$ values, beginning with the $i$th point.

Second, given a vector $X_i^m$, the similarity degree $D_{ij}^m$ between $X_i^m$ and $X_j^m$ is defined by the fuzzy membership function:

$$D_{ij}^m = fu(d_{ij}^m, n, r) = exp\left(-\frac{(d_{ij}^m)^n}{r}\right) \quad (7)$$

where the fuzzy membership function $fu$ is an exponential function that is more appropriate for processing physiological signals in term of robustness to noise and independence on the data length [25, 31, 32], and $d_{ij}^m$ is the maximum absolute difference between the corresponding scalar components of $X_i^m$ and $X_j^m$.

The parameter $m$ is the length of the sequences to be compared to other entropy algorithms. The other two parameters, $r$ and $n$, determine the width and the gradient of the boundary of the fuzzy membership function, respectively.

Then, the function $\varphi^m$ is constructed. Similarly, for $m + 1$, the above steps are repeated and denoted $\varphi^{m+1}(n, r)$.

$$\varphi^m(n, r) = (N-m)^{-1} \sum_{i=1}^{N-m} \left((N-m-1)^{-1} \sum_{j=1, j \neq i}^{N-m} D_{ij}^m\right) \quad (8)$$

Finally, the $entropy\,(m, n, r, N)$ parameter of the sequence $\{\hat{x}(i): 1 \leq i \leq N\}$ is defined as the negative natural logarithm of the deviation of $\varphi^m$ from $\varphi^{m+1}$:

$$entropy(m, n, r, N) = \ln \varphi^m(n, r) - \ln \varphi^{m+1}(n, r) \quad (9)$$

Of note, the measurement of Eq. (9) is consistent with the previously developed approximate entropy, sample entropy, and fuzzy entropy. The entropy measures the likelihood that runs of the patterns that are close to the observations remain close on the next incremental comparisons, which can quantify physiological time-series complexity. Please see below the proof to verify that the measurement in Eq. (9) represents an entropy.

*Proof Eq. (9):*

Let $B_i$ be the number of vectors $X_i^m$ within $r$ of $X_j^m$ and Let $A_i$ be the number of vectors $X_i^{m+1}$ within $r$ of $X_j^{m+1}$.

Define the function $C_i^m(n, r) = (B_i)/(N - m)$ and $C_i^{m+1}(n, r) = (A_i)/(N - m + 1)$.

$C_i^m(n,r)$ is the probability that any vector $X_i^m$ is within $r$ of $X_j^m$.

$C_i^{m+1}(n,r)$ is the probability that any vector $X_i^{m+1}$ is within $r$ of $X_j^{m+1}$.

Then, based on the Grassberger and Procaccia's work [33], define the function:

$$\varphi^m(n,r) = (N-m)^{-1} \sum_{i=1}^{N-m} In\ [C_i^m(r)]$$

which is the average of natural logarithms of functions $C_i^m(n,r)$

Similarly,

$$\varphi^{m+1}(n,r) = (N-m+1)^{-1} \sum_{i=1}^{N-m+1} In\ [C_i^{m+1}(n,r)]$$

With the limitation as $N \to \infty$,

$$\varphi^m(n,r) = (N-m)^{-1} \sum_{i=1}^{N-m} In\ [C_i^m(r)]$$

$$\varphi^{m+1}(n,r) = (N-m)^{-1} \sum_{i=1}^{N-m} In\ [C_i^{m+1}(r)]$$

Based on Pincus's work [22], entropy measures the likelihood that runs of the patterns that are close to $m$ observations remain close on the next incremental comparisons.

$$Entropy = \ln \varphi^m(n,r) - \ln \varphi^{m+1}(n,r)$$
$$= (N-m)^{-1} \{ \sum_{i=1}^{N-m} In\ [C_i^m(n,r) - InC_i^{m+1}(n,r)] \}$$

which equals the average over $i$ of $In\ [C_i^m(n,r)/C_i^{m+1}(n,r)] = In\ [B_i/A_i]$.

We noted that the ratio $[B_i/A_i]$ represents an entropy.

*D. Relative inherent fuzzy entropy*

To minimize individual differences, we introduced a relative multi-scale inherent fuzzy entropy ($RE$). First, the inherent fuzzy entropy at baseline (resting) and during stimulus sessions is calculated. The two sessions are simplified to $Entropy_{baseline}$ and $Entropy_{SSVEP}$, respectively.

Next, we calculated the variation in the multi-scale inherent fuzzy entropy between the baseline (resting) condition and the stimulus condition during which SSVEPs were induced in five stimulus trials. This function is termed the relative inherent fuzzy entropy, which is expressed as:

$$RE_k = Entropy_{SSVEP(k)} - Entropy_{baseline} \quad (10)$$

where $k$ is the stimulus time.

## III. EXPERIMENTAL PROCEDURES AND METHODS

*A. Participants*

Forty outpatients with migraine without aura (F:M = 30:10, mean age: 38.1 ± 8.2) were recruited from the Headache Clinic of Taipei Veterans General Hospital and were asked to keep a headache diary to determine migraine phases on a daily basis. All enrolled patients fulfilled the diagnostic criteria of the International Classification of Headache Disorders 2nd edition (ICHD-II) and had a migraine frequency ranging from 1 to 6 days per month. Forty age- and sex-matched HCs (F:M = 32:8, mean age: 36.1 ± 9.8) were recruited from hospital colleagues or their relatives or friends. The individuals who served as HCs did not have a past medical history or a family history of migraine. As we used the same dataset in another recent study [29], the comparisons of demographics, headache profile, and psychological characteristics between HCs and patients are summarized in that report.

As shown in Fig. 2, the days on which the EEG examinations were performed were classified into one of four migraine phases (inter-ictal, pre-ictal, ictal, or post-ictal) based on the headache diary. The ictal phase was coded when the patient was suffering from a migraine attack on the day of the EEG study. Based on previous criteria [34, 35], the pre-ictal and post-ictal phases were coded on the day of the EEG study if the patient was within 72 hours before or after an ictal phase, respectively. The inter-ictal phase was coded if the patient had not had a migraine attack within 72 hours before or after the EEG examination. Only EEG data collected during the inter-ictal and pre-ictal phases were selected for analysis in this study, as these periods are significant phases before migraine attacks and potential to detect the transition of migraine phases.

All the participants had normal vision and no systemic diseases, connective tissue disorders, neurological, psychiatric disorders, or other painful conditions according to their self-reports. None of our migraine patients received preventive treatment, and they were asked not to take any analgesics within two days before the EEG recording. The Institutional Review Board of Taipei Veterans General Hospital approved

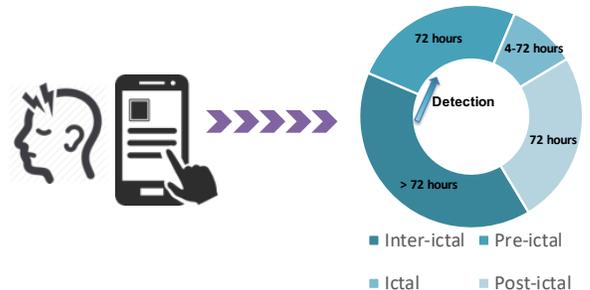

**Figure 2** Headache diary and migraine cycle.

this study. Informed consent was obtained from all participants before they entered in the study.

*B. Wearable EEG device*

EEG signals were recorded at a sampling rate of 500 Hz by a "Mindo" EEG device (Brain Rhythm Inc., Zhubei District, Hsinchu, Taiwan), which is a wearable headband EEG device with dry sensors [36]. Each dry-contact electrode was designed to include a probe head, a plunger, a spring, and a barrel. The probes were inserted into a flexible substrate via an established injection molding procedure using a one-time forming process. These dry electrodes are more convenient for measuring EEG signals than conventional wet electrodes and are preferred because they avoid the need to use conductive gel and extensive skin preparation procedures while achieving a signal quality comparable to that of wet electrodes. In this study, as shown in Fig. 3-A, four dry-contact electrodes (Fpz, O1, Oz, and O2) were placed according to the extended International 10–20 system, and two extra channels (A1 and A2) were used as reference channels. The collected EEG signals were transmitted by Bluetooth to a personal computer.

*C. Experimental paradigm*

The EEG experiment was performed in a static room at Taipei Veterans General Hospital, Taiwan. To avoid light source interference, we turned the fluorescent lamps off during the experimental procedure. As shown in Fig. 3-B, the illumination of monitor (Viewsonic V3D231) was 90 lux s/pulse at 20 cm in front of the participants' eyes, and the participants were asked to put their chin on the shelf. Additionally, the screen offers the functions to inspect the raw EEG signals and stimulate repetitive visual stimulus flicks in the form of alternating graphical patterns.

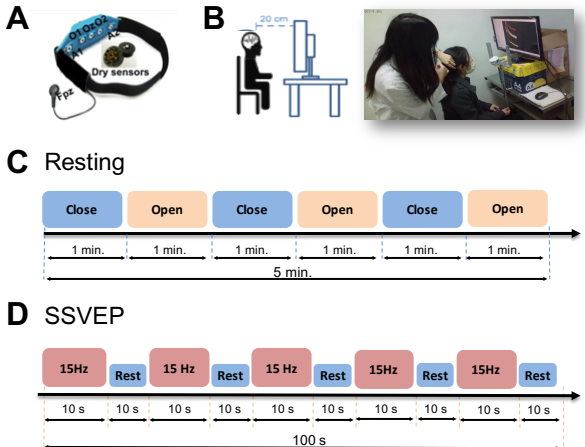

**Figure 3** Experimental paradigm. (A) Wearable headband EEG device; (B) experimental environment and setting; (C) experiment during resting condition; (D) experiment during SSVEP condition.

As shown in Fig. 3-C and Fig. 3-D, the experiment consisted of the following two sessions: a resting session and a visual stimulus session with five stimulus trials. The resting session comprised three epochs, during which the eyes were open for 1 min and closed for 1 min. To prevent the migraine patients from receiving excessive visual stimuli, we administered five trials of 10-second, 15-Hz stimuli to evoke SSVEPs with 10-second intervals between stimuli in the eyes-closed condition. To evaluate the EEG complexity changes related to visual stimulation, resting EEG recording during eye-closed condition was served as the baseline [37] and was then extracted to calculate the relative multi-scale inherent fuzzy entropy.

Of note, patients and HCs participated in five sessions of identical resting-state and SSVEP EEG examinations over 3–7 months, and each examination was separated by an interval of 2-8 weeks. This protocol aimed to acquire EEG data for patients who experienced different phases of migraine, particularly the inter-ictal and pre-ictal phases.

*D. EEG processing*

All the EEG data were analyzed using the EEGLAB toolbox (http://sccn.ucsd.edu/eeglab/) with MATLAB software (Mathworks, Inc.), an open source toolbox for electrophysiological signal processing. The EEG signals were recorded from the O1, Oz, O2 and Fpz electrodes for the resting and SSVEP epochs during the eyes-closed condition.

1. Pre-processing

The original EEG was reviewed by experienced EEG specialists. EEG activity was sampled at 500 Hz, down-sampled to 250 Hz, and then filtered through 1-Hz high-pass and 30-Hz low-pass finite impulse-response filters. Segments contaminated with non-physiological artifacts, including movement artifacts, electrode detachment, sweating artifacts, or 60-Hz noise were marked and discarded.

2. CCA-based artificial removal

Several studies have applied canonical correlation analysis (CCA) to improve signal quality [38-40], and we also have reported that the quality of EEG signals can be enhanced by CCA-based algorithms [41]. In this study, the filtered EEG signal $s(x)$ was expressed as:

$$\text{EEG data: } S(x) = \begin{bmatrix} O1 \\ Oz \\ O2 \\ Fpz \end{bmatrix} \qquad (11)$$

The template sinusoidal and cosinusoidal signals associated with the flickering frequency $S(y)$ were expressed as:

Template signal: $S(y) = \begin{bmatrix} \sin(2\pi f_1 n) \\ \cos(2\pi f_1 n) \\ \sin(2\pi f_2 n) \\ \cos(2\pi f_2 n) \end{bmatrix}$ (12)

where $f_1$ is the frequency of the stimulation, $f_2$ is the second harmonic frequency to the stimulus frequency, and $n$ is the length of the EEG signal according to the following equation:

$n = \frac{1}{f_s}$ (13)

where $f_s$ is the sampling rate.

The template sine and cosine signals were 15 Hz, and the template second harmonic sine and cosine signals were 30 Hz.

Then, we calculated the correlation between $S(x)$ and $S(y)$:

$[C_{xx}^{-1} C_{xy} C_{yy}^{-1} C_{yx}] W_x = \rho^2 W_x$ (14)

where $C_{xx}$ is the covariance matrix of $S(x)$, and $C_{yy}$ is the covariance matrix of $S(y)$. $C_{xy}$ and $C_{yx}$ are the cross-covariance matrices between $S(x)$ and $S(y)$. $W_x$ contains the eigenvectors and eigenvalues of matrices $S(x)$ and $S(y)$.

Considering the components of the eigenvector that are similar to the stimulus frequency, the first two selected components were comprised matrix $A$. In addition, we transposed matrix $A$ and multiplied it by the original EEG data $S(x)$:

$Y' = A^T S(x)$ (15)

The inverse of matrix $A$ was also transposed and multiplied by matrix $Y'$:

$S(x)' = (A^T)^{-1} Y'$ (16)

The matrix $S(x)'$ represents the EEG signals after CCA-based artifact removal.

Finally, the artifact-free EEG signals were inspected again using the automatic continuous rejection function in EEGLAB.

3. Entropy estimation

The artifact-free EEG data were estimated by the relative multi-scale inherent fuzzy entropy algorithm (refer to Section II). During the multi-scale procedure, each coarse-grained time series was calculated for different temporal scales from 1 to 20. Of note, the values of the relative inherent fuzzy entropy in the resting and stimulus conditions were obtained by averaging the entropy estimates of three 1-min eyes-closed blocks and 10-s SSVEP blocks, respectively. The prefrontal entropy was calculated from the entropy values calculated from data recorded from the Fpz electrode, and the occipital entropy was calculated by averaging the entropy values corresponding to the O1, Oz, O2 electrodes. The entropy estimates for each patient were averaged over the same migraine phase if the participant had more than one EEG recordings during inter-ictal or pre-ictal phase, and the entropy estimates of each HC were averaged over the examinations.

### E. Classification models

A binary classification model was developed to discriminate among migraine phases (inter-ictal vs. pre-ictal). In this study, we employed six commonly used algorithms, including linear discriminant analysis (LDA), a *k*-nearest neighbors classifier (*k*NN), multilayer perceptron (MLP), a Bayesian classifier, a support vector machine (SVM) with a linear or radial basis function (RBF) kernel [42], Gaussian processes (GP) [43], random forest (RF) [44], and adaptive boosting (AdaBoost) [45] to classify migraine phases based on EEG entropy features with significant levels. These classification algorithms were all implemented using PRTools [46], LIBSVM [47], GPML [43] or MATLAB File Exchanges in MATLAB software.

The performances of the six algorithms were validated and compared via a 3-fold cross-validation procedure. That is, all data were randomly partitioned into 3 approximately equally-sized clusters. Two clusters were combined and used as the training data, and the remaining cluster was retained as the validation data to test the model. For example, the original sample (40 patients) was randomly partitioned into 3 equally-sized subsamples. Of the 3 subsamples, 2 subsamples (~27 patients) were used as training data, and the remaining subsample (~13 patients) was retained as the validation data for testing the model. The cross-validation process was repeated 100 times. Additionally, due to the limitation of a small dataset, the approach for optimizing hyperparameter is tuned by cross-validation (*k*-fold) to evaluate the model performance. Specifically, we first split train and test subsets with 3-folds, and randomized search of parameters on train subset. Then, we selected the best estimator obtained after the randomized search based on accuracy. Finally, we tested the tuned classifiers with test subset and obtained the accuracy scores.

The performance metrics covered classification accuracy, recall, precision, and F-measure [48]. Specifically, accuracy is the most intuitive performance measure, which is simply a ratio of the correctly predicted observations to the total observations. Recall (also called sensitivity) is the ratio of correctly predicted positive observations to all observations in the actual class. Precision (also called positive predictive value) is the ratio of correctly predicted positive observations to the total predicted positive observations. F-measure is the weighted average of precision and recall, which takes both false positives and false negatives into account. Additionally, the receiver operating characteristic (ROC) curves and area under the curve (AUC) were employed to evaluate the performance of various classifiers.

## F. Statistical analysis

To determine the independence of different entropy variables in of repetitive visual stimulation (i.e. between the 1st and 5th stimuli), we performed paired *t*-tests for each entropy scale to compare EEG complexity within the same group (HCs, inter-ictal, and pre-ictal patients). Furthermore, paired *t*-tests were applied to compare EEG complexity between inter-ictal and pre-ictal patients, and independent *t*-tests were used to test for differences between groups (HCs vs. inter-ictal or pre-ictal patients). False discovery rate (FDR) correction was used to control for multiple comparisons. The intra-class correlation was estimated to quantify a test-retest reliability. One-way ANOVA was used to compare the performances of the classification algorithms, followed by Tukey's post hoc test to test all pairwise comparisons. All the statistical tests were two-tailed, and statistical significance was set at $p < 0.05$.

## IV. RESULTS

EEG complexity was measured with multi-scale relative inherent fuzzy entropy (RE), simply referred to as "entropy" in the following sections, over different time scales τ ranging from 1 to 20 in repetitive SSVEPs from prefrontal and occipital areas. With increased stimulus times, we noted monotonic enhancement in the occipital EEG entropy of HCs and patients during the inter-ictal phase, but a monotonic reduction in the occipital EEG entropy in patients during the pre-ictal phase (Fig. 4-A). However, no significant differences or trends in the EEG entropy were observed among the different groups in the prefrontal region. Therefore, in the following sections, we present the detailed findings of the occipital regions.

## A. Intra-group comparisons of EEG entropy

In this section, we compared the differences in EEG entropy between different sessions (1st stimuli vs. 5th stimuli) in three groups of participants, i.e., HCs, inter-ictal patients, and pre-ictal patients. As shown in Fig. 4-B, EEG entropy showed a decreasing trend in entropy with an increasing time scale in the 1st stimulus session, which changed to an increasing trend in the 5th stimulus session in HCs. Furthermore, when comparing EEG entropy between the 1st and 5th stimulus sessions, the results showed an enhancement in occipital EEG entropy with increasing stimulus times for HCs. Table I shows the values of EEG entropy and *p* values in the 1st and 5th stimuli sessions. The EEG entropy increased from negative to positive values in the HCs group. The paired *t*-tests revealed that the EEG entropy in the 5th SSVEP session was significantly higher than that in the 1st SSVEP session in most time scales (FDR-adjusted $p < 0.05$).

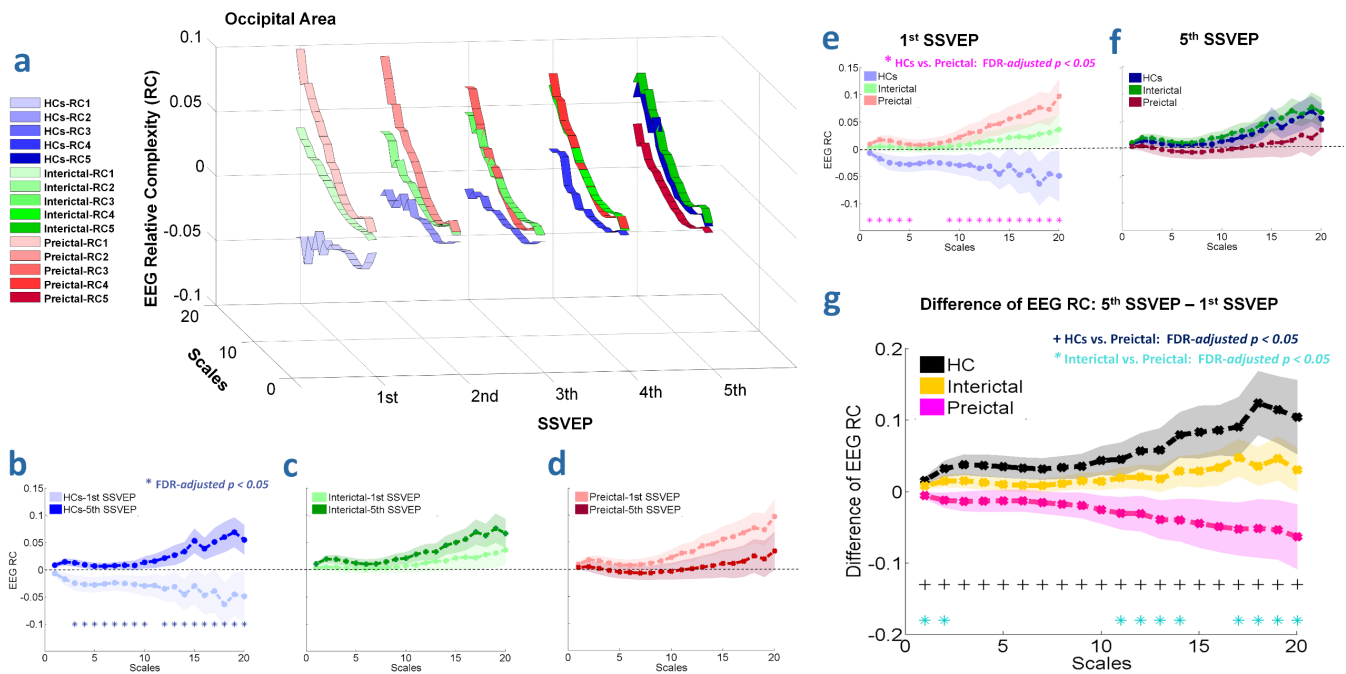

**Figure 4** The trends in EEG entropy in the occipital area. (A) The changes in EEG entropy during the five (1st, 2nd, 3rd, 4th, and 5th marks) SSVEP stimuli over time scales ranging from 1 to 20 in HCs and migraine patients during the inter-ictal and pre-ictal phases. (B) Comparisons of EEG entropy between the 1st and 5th stimuli over time scales ranging from 1 to 20 in HCs. (C) Comparison of EEG entropy between the 1st and 5th stimuli over a time scale of 1 to 20 in patients of inter-ictal phase. (D) Comparisons of EEG entropy between the 1st and 5th stimuli over time scales ranging from 1 to 20 in patients during the pre-ictal phase. (E) Comparisons of EEG entropy among the inter-ictal phase, the pre-ictal phase and HCs over time scales ranging from 1 to 20 in the 1st stimuli session. (F) Comparisons of EEG entropy among the inter-ictal phase, the pre-ictal phase and HCs over time scales ranging from 1 to 20 in the 5th stimuli session. (G) Variance (5th stimuli EEG entropy minus 1st stimuli EEG entropy) comparisons among the inter-ictal phase, the pre-ictal phase and HCs. Of note, the traces represent the mean ± standard deviation (SD) of the EEG entropy of the inter-ictal phase, the pre-ictal phase or HCs. The asterisk or cross denotes a significant difference between different conditions (FDR-adjusted $p < 0.05$).

For migraine patients in the inter-ictal phase (Fig. 4-C), EEG entropy had an ascending tendency with increasing time scales in the 1st and 5th stimulus sessions. Furthermore, EEG entropy had an increasing trend from the 1st to the 5th stimulus in all time scales, although the difference was not significant. However, for migraine patients in the pre-ictal phase (Fig. 4-D), although EEG entropy also retained an ascending tendency with increasing time scales in the 1st and 5th stimulus sessions, the entropy showed an opposite trend from the 1st to the 5th stimulus in all time scales. That is, EEG entropy showed a decreasing trend in the 5th stimuli session compared with the 1st stimuli session.

*B. Inter-group comparisons of EEG entropy*

In this section, we compared the differences in EEG entropy in the specific stimuli sessions among the three groups (HCs, inter-ictal patients and pre-ictal patients). During the 1st stimuli session (Fig. 4-E), the HCs group presented lower occipital EEG entropy relative to the migraine patients. The EEG entropy values of the HCs and the migraine patients were negative and positive, respectively, for all time scales. We did not find a significant difference in EEG entropy between the HCs and the patients in the inter-ictal phase, but patients in the pre-ictal phase had significantly higher EEG entropy compared with the HCs (FDR-adjusted $p < 0.05$). In Table I, we present the values of EEG entropy and the $p$ values for HCs and patients in the pre-ictal phase. In migraine patients, the pre-ictal phase was associated with a rising trend in EEG entropy at large time scales compared with the inter-ictal phase, although there were no significant differences between these two phases.

By the 5th stimulus (Fig. 4-F), participants in all three groups exhibited positive entropy values in most time scales. Patients in the pre-ictal phase presented the lowest occipital EEG entropy relative to HCs and patients in the inter-ictal phase, although the EEG entropy did not significantly differ among the three groups. Of note, EEG entropy of HCs is similar to that of patients of inter-ictal phase in most time scales.

Additionally, we defined the transitional variance of EEG entropy as the difference in EEG entropy between the 1st and 5th stimulus sessions. As shown in Fig. 4-G, patients in the pre-ictal phase exhibited a significantly lower transitional variance of EEG entropy than that in the inter-ictal phase for most large time scales. Similarly, patients in the pre-ictal phase had a significantly lower transitional variance of EEG entropy than the HCs for all time scales (FDR-adjusted $p < 0.05$). Table II shows the transitional variance of EEG entropy and the $p$ values for HCs and patients in the inter-ictal and pre-ictal phases. Considering the transitional variance of EEG entropy in individuals from the inter-ictal to pre-ictal phase, for the time scale of 20, entropy decreased in 28 of the 40 patients (70%) when they entered the pre-ictal phase from the inter-ictal phase. In contrast, an increment in entropy was observed in 12 patients. Additionally, 8 patients with two pre-ictal examinations were selected for the test-retest reliability. According to the quoted guidelines for interpretation of inter-rater agreement measures, our results showed a good reliability with an intra-class correlation

**Table I** EEG entropy (mean (standard deviation)) in HCs and patients.

| $\tau$ | HC | | Pre-ictal | $P^1$ | $P^2$ |
| --- | --- | --- | --- | --- | --- |
| | *1st SSVEP* | *5th SSVEP* | *1st SSVEP* | | |
| 1 | -0.007 (0.005) | 0.008 (0.003) | 0.009 (0.003) | 0.053 | 0.014 |
| 2 | -0.018 (0.011) | 0.014 (0.006) | 0.017 (0.007) | 0.059 | 0.012 |
| 3 | -0.025 (0.013) | 0.012 (0.006) | 0.015 (0.009) | 0.014 | 0.016 |
| 4 | -0.027 (0.014) | 0.009 (0.006) | 0.011 (0.009) | 0.025 | 0.028 |
| 5 | -0.028 (0.014) | 0.006 (0.006) | 0.008 (0.010) | 0.032 | 0.039 |
| 6 | -0.026 (0.014) | 0.006 (0.006) | 0.007 (0.010) | 0.037 | 0.057 |
| 7 | -0.024 (0.014) | 0.007 (0.006) | 0.008 (0.010) | 0.045 | 0.069 |
| 8 | -0.026 (0.014) | 0.008 (0.007) | 0.011 (0.011) | 0.043 | 0.057 |
| 9 | -0.027 (0.016) | 0.008 (0.009) | 0.015 (0.012) | 0.048 | 0.041 |
| 10 | -0.030 (0.018) | 0.013 (0.010) | 0.021 (0.014) | 0.039 | 0.029 |
| 11 | -0.029 (0.020) | 0.015 (0.013) | 0.029 (0.016) | 0.064 | 0.029 |
| 12 | -0.035 (0.023) | 0.021 (0.014) | 0.032 (0.018) | 0.030 | 0.024 |
| 13 | -0.032 (0.026) | 0.026 (0.018) | 0.043 (0.019) | 0.045 | 0.027 |
| 14 | -0.046 (0.028) | 0.033 (0.017) | 0.046 (0.021) | 0.018 | 0.012 |
| 15 | -0.030 (0.033) | 0.053 (0.022) | 0.055 (0.023) | 0.030 | 0.039 |
| 16 | -0.047 (0.035) | 0.038 (0.020) | 0.060 (0.024) | 0.019 | 0.014 |
| 17 | -0.039 (0.039) | 0.051 (0.025) | 0.067 (0.026) | 0.036 | 0.027 |
| 18 | -0.064 (0.039) | 0.060 (0.027) | 0.076 (0.028) | 0.008 | 0.005 |
| 19 | -0.046 (0.039) | 0.068 (0.027) | 0.072 (0.029) | 0.019 | 0.019 |
| 20 | -0.049 (0.047) | 0.054 (0.027) | 0.097 (0.032) | 0.049 | 0.012 |

$\tau$ represents the time scales.
[1] 1st SSVEP vs. 5th SSVEP for HCs.
[2] Pre-ictal patients vs. HCs in the 1st SSVEP session.

**Table II** Variance of EEG entropy (mean (standard deviation)) in HCs and patients

| $\tau$ | HC | Patients | | $P^1$ | $P^2$ |
| --- | --- | --- | --- | --- | --- |
| | | *Inter-ictal* | *Pre-ictal* | | |
| 1 | 0.015 (0.005) | 0.007 (0.004) | -0.005 (0.004) | 0.004 | 0.024 |
| 2 | 0.031 (0.011) | 0.014 (0.008) | -0.012 (0.010) | 0.006 | 0.035 |
| 3 | 0.037 (0.014) | 0.014 (0.010) | -0.014 (0.014) | 0.013 | 0.074 |
| 4 | 0.036 (0.015) | 0.012 (0.010) | -0.013 (0.015) | 0.024 | 0.129 |
| 5 | 0.034 (0.015) | 0.009 (0.009) | -0.013 (0.015) | 0.030 | 0.182 |
| 6 | 0.032 (0.015) | 0.008 (0.009) | -0.013 (0.015) | 0.036 | 0.208 |
| 7 | 0.031 (0.015) | 0.008 (0.009) | -0.015 (0.016) | 0.034 | 0.183 |
| 8 | 0.033 (0.016) | 0.010 (0.010) | -0.017 (0.018) | 0.028 | 0.125 |
| 9 | 0.034 (0.017) | 0.014 (0.011) | -0.019 (0.019) | 0.033 | 0.095 |
| 10 | 0.043 (0.020) | 0.014 (0.012) | -0.025 (0.022) | 0.016 | 0.081 |
| 11 | 0.044 (0.023) | 0.019 (0.014) | -0.030 (0.024) | 0.023 | 0.047 |
| 12 | 0.056 (0.025) | 0.020 (0.010) | -0.031 (0.027) | 0.015 | 0.049 |
| 13 | 0.057 (0.030) | 0.017 (0.018) | -0.039 (0.029) | 0.019 | 0.045 |
| 14 | 0.078 (0.032) | 0.028 (0.021) | -0.040 (0.034) | 0.008 | 0.048 |
| 15 | 0.082 (0.036) | 0.028 (0.023) | -0.044 (0.035) | 0.013 | 0.070 |
| 16 | 0.085 (0.035) | 0.033 (0.025) | -0.049 (0.037) | 0.008 | 0.050 |
| 17 | 0.090 (0.041) | 0.047 (0.026) | -0.052 (0.040) | 0.012 | 0.019 |
| 18 | 0.123 (0.044) | 0.034 (0.027) | -0.051 (0.041) | 0.005 | 0.044 |
| 19 | 0.113 (0.046) | 0.045 (0.030) | -0.053 (0.041) | 0.009 | 0.036 |
| 20 | 0.103 (0.052) | 0.029 (0.031) | -0.063 (0.045) | 0.018 | 0.049 |

$\tau$ represents the time scales.
[1] Pre-ictal patients vs. HCs.
[2] Pre-ictal vs. inter-ictal phases.

coefficient $r_1$ of 0.70 ($p = 0.04$).

## C. Performance of classification models

To capture useful input variables, we selected features (transitional variance of EEG entropy) for specific time scales with significant changes. Thus, the transitional variance of EEG entropy with time scales of $\tau = 1, 2, 11, 12, 13, 14, 17, 18, 19, 20$ were chosen as the input features of the classification models.

As shown in Table III, the performances of six binary classification models (LDA, kNN, MLP, Bayesian, Linear/RBF-SVM, GP, RF, and AdaBoost) for classifying migraine phases (inter-ictal vs. pre-ictal) were evaluated. Furthermore, we selected four matrices (accuracy, recall, precision, and F-measure) to assess classification performance. In terms of accuracy and recall, the AdaBoost ensemble learning classifier (81±6% accuracy and 80±6% recall) significantly outperformed other classifiers ($p < 0.05$) except for the RBF-SVM. In terms of precision and F-measure, the AdaBoost ensemble learning classifier (79±6% precision and 78±5% F-measure) had comparable results that were significantly better than the results obtained with the LDA, kNN, MLP, Bayesian, Linear-SVM, GP and RF classifiers ($p < 0.05$).

Additionally, the ROC curves and AUC were employed to evaluate the performance of various classifiers. Figure 5 showed the ROC curves and AUC of the classifiers to compare the classification performance among various classifiers. Of note, AdaBoost had the highest AUC of 0.87, and SVMRBF had the second-highest AUC of 0.84 in estimating the migraine phases.

**Table III** Performances of the classification models

| | | Accuracy | Recall | Precision | F-measure |
|---|---|---|---|---|---|
| | LDA | 0.72 (0.05)* | 0.69 (0.05)* | 0.72 (0.05)* | 0.71 (0.05)* |
| | kNN | 0.70 (0.06)* | 0.68 (0.06)* | 0.70 (0.06)* | 0.69 (0.06)* |
| | MLP | 0.69 (0.07)* | 0.63 (0.05)* | 0.67 (0.06)* | 0.66 (0.07)* |
| | Bayesian | 0.67 (0.06)* | 0.63 (0.06)* | 0.67 (0.06)* | 0.67 (0.06)* |
| S V M | Linear | 0.74 (0.05)* | 0.71 (0.05)* | 0.75 (0.06)* | 0.73 (0.05)* |
| | RBF | 0.79 (0.05) | 0.78 (0.06) | 0.77 (0.05) | 0.76 (0.05) |
| | GP | 0.73 (0.05)* | 0.70 (0.05)* | 0.72 (0.05)* | 0.72 (0.05)* |
| | RF | 0.75 (0.06)* | 0.73 (0.05)* | 0.75 (0.06)* | 0.74 (0.05)* |
| | **AdaBoost** | **0.81 (0.06)** | **0.80 (0.06)** | **0.79 (0.06)** | **0.78 (0.06)** |

Parameters: LDA: default settings; KNN: $k$=3; MLP: structure with one hidden layer and number of units=5; Bayesian: default settings; SVM with linear kernel: $c$=0; SVM with RBF kernel: $c$=15, $g$=10; GP: exponential covariance [-0.73 -0.12] and Gaussian likelihood -2.58; RF: the number of decision trees $n$=20; AdaBoost: the number of weak classifiers $n$=8 (LDA, kNN, MLP, Bayesian, Linear-SVM, RBF-SVM, GP, and RF).

\*: After utilizing one-way ANOVA, Tukey's post hoc test was performed for pair-wise comparisons: $p < 0.05$ (AdaBoost vs. Other Classifiers).

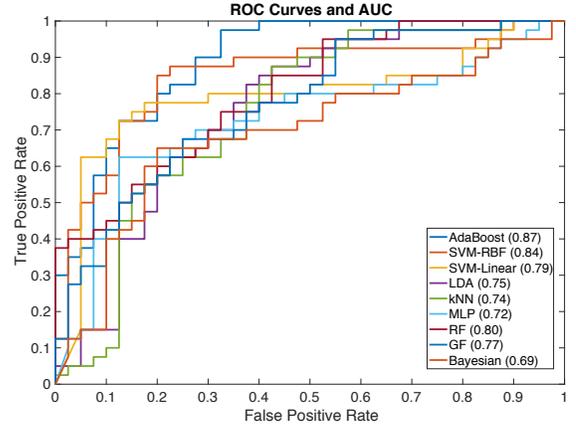

**Figure 5** ROC curves and AUC are analyzed and plotted for each of the classifiers. The AUC is marked in the brackets of the legends (e.g., 0.87 AUC in AdaBoost).

## D. Performance of competing entropy algorithms

In this section, we used the same EEG processing steps but estimated entropy using the competing entropy algorithms (ApEn, SampEn, and FuzzEn). In HCs (Fig. 6), we could not distinguish EEG entropy between the 1st and 5th stimuli sessions using ApEn or SampEn without a fuzzy structure. However, the FuzzEn algorithm could only distinguish the EEG entropy of these two sessions at larger time scales ($\tau = 17, 18, 19, 20$), suggesting that the fuzzy structure (including FuzzEn and Inherent FuzzEn) is better than the non-fuzzy structure (including ApEn and SampEn) in estimating EEG entropy in SSVEPs.

Additionally, there was no significant difference in EEG entropy between each session in migraine patients, indicating that the relative inherent fuzzy entropy algorithm is superior to the other competing models, including ApEn, SampEn and FuzzEn, for use in SSVEP-based migraine studies.

## V. DISCUSSION

### A. Complexity characteristics

Systemic analyses indicate that entropy dynamics reflect the nonlinear complex characteristics of the brain that allow it to adapt to constantly changing stimulus situations rather than the linear characteristics of the brain [49]. The associated entropy models, which can be used to quantify brain complexity to express the robustness of brain systems, are crucial for quantifying the critical characteristics of nonlinear neuro-dynamics [50]. Additionally, as biological systems functioning on different time scales may exhibit different behaviors, so bio-signals are often examined at multiple time scales. Our results showed that EEG complexity was enhanced in the occipital area after repetitive visual stimulations at different time scales, which was in line with the finding of a previous study showing that EEG entropy exhibited an increasing trend in response to long-term audio-visual stimulation [51].

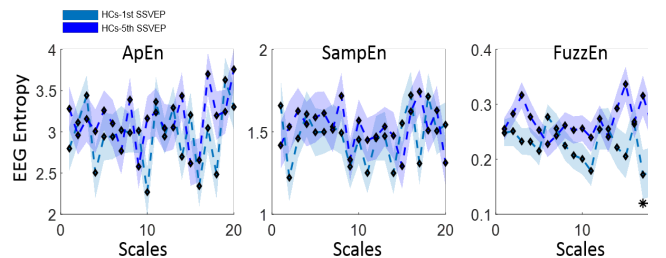

**Figure 6** Comparisons of EEG entropy between the 1st and 5th stimuli over 1 to 20 time scales in HCs as measured by ApEn, SampEn and FuzzEn.

On the other hand, healthy individuals learn to stop responding to stimuli that are no longer biologically relevant [52, 53]. For example, healthy humans can habituate to repeated visual stimuli when they learn these have no consequences. This habituation performance is a form of adaptive behavior that reflects the robustness of brain systems. Our finding that EEG complexity was significantly enhanced during repetitive SSVEPs in HCs is considered another form of habitation indicating the robustness of brain systems. Traditionally researchers who study habituation have focused on a time/frequency dimension of the behavior (i.e. response potential or magnitude). Our high throughput entropy analyses of habituation for complexity estimation have changed this view. Our study offers a new way to think about the role of inherent fuzzy entropy in the context of adaptive behavioral strategies, i.e., habituation.

*B. EEG complexity in migraine patients*

Prior studies have noted that healthy brain system present robustness with high complexity but diseased brain systems show vulnerable characteristics with low complexity [20, 29]. Healthy subjects exhibit normal habituation to repetitive visual stimulation [30] but dishabituation has been observed in migraine patients [6-11]. Our study discovered changes in EEG complexity in migraine patients by using a multi-scale relative inherent fuzzy entropy algorithm in a repetitive SSVEP environment. We found that occipital complexity was slightly, although not significant, enhanced with increasing stimulus times for migraine patients in the inter-ictal phase, but a reversed trend was observed for migraine patients in the pre-ictal phase. Thus, our findings are consistent with the prior studies and may suggest that migraine patients in the pre-ictal phase are more vulnerable to adapt a repetitive visual simulation environment than those in the inter-ictal phase.

Regarding the values of relative inherent fuzzy entropy in the 1$^{st}$ SSVEP session, a key result is that migraine patients and HCs have positive and negative entropy values, respectively. This finding suggests that migraine patients and HCs have higher and lower EEG complexities, respectively, relative to baseline (resting-state) when responding to the 1$^{st}$ visual stimulation.

Additionally, it would be, nevertheless, of great interest to evaluate the EEG complexity by using different stimulation frequencies, either lower or higher ones. Since previous SSVEP studies all demonstrated a hyper-responsive pattern in patients with migraine throughout a low- to median-to-high frequency range, we speculate the results would be similar. Further studies are required to confirm this point.

*C. EEG-based headband BCI migraine system*

Because of its broad availability and cost-effectiveness, EEG is widely used as a non-invasive means to assess dynamic changes in brain electrical activity. The rapid development of dry sensors and wearable devices [36, 54, 55] has led to a reduction in the preparatory work required for long-term monitoring. Moreover, the headband design with occipital electrodes is convenient for long-term monitoring and daily use [36, 54, 55]. Therefore, it is possible to implement EEG-based models in laboratories and real-world settings.

Considering the transitional variance in EEG entropy, our findings showed that migraine patients in the pre-ictal phase presented significantly lower occipital complexity than migraine patients in the inter-ictal phase or HCs. Due to the distinctive specifications of each classifier, we used 9 EEG-based classifiers with the input of the transitional variance in EEG entropy, to test and compare the performance on which of two migraine phases a new EEG signal belongs. Of them, the SVM constructs a hyperplane in a high-dimensional space, which has the largest distance to the nearest training-data point of any class. The GF is a probabilistic classification model specified by a mean function and a covariance function. The RF is basically an ensemble of decision trees, and each tree classifies the dataset using a subset of variables. Especially, the AdaBoost, a machine learning meta-algorithm, can be used in conjunction with many other types of classifiers to improve the performance. In the study, the output of the other classifiers including LDA, kNN, MLP, Bayesian, Linear-SVM, RBF-SVM, GP, and RF, called weak classifiers, is combined into a weighted sum that represents the final output of the boosted classifier. This outstanding outcome contributed to discriminating between inter-ictal and pre-ictal migraine phases with 81±6% accuracy, 80±6% recall, 79±6% precision and 78±6% F-measure, as well as the AUC of 0.87.

In a 2-class decoding problem with a small set ($n$=80), the classification performance significantly exceeds chance if the accuracy reaches 70% ($p < 10^{-4}$) [56]. In our study, the accuracy of classifiers is larger than 70%, which indicated that the performance of our classifiers exceeded the chance level. Thus, the wearable EEG solution and the characteristics of multi-scale relative inherent fuzzy entropy can be easily assessed in repetitive visual stimulation paradigms. However, the transitional variance of EEG entropy indeed affects the performance of classifiers to obtain a good generalization of validation. What is more, the computational time of the multi-scale inherent fuzzy entropy algorithm prevents online applications or even the processing of long data sets, due to the computational time of the fuzzy entropy value. Our next step will conduct a follow-up study to record daily EEG signals in out-patient individuals to evaluate the transitional variance between the two migraine phases, and calculate entropy values with less computational time. In the future, we believe that occipital complexity features can be evaluated to develop a brain-computer interface (BCI) system that can be used to recognize migraine phases before a migraine attack.

*D. Advantages of inherent fuzzy entropy*

Despite recent progress that has been made in visual stimulus research, extracting corticocerebral complexity using EEG-based multi-scale entropy approaches that can be used to determine the robustness of brain systems remains challenging.

In previous studies, researchers have generally ignored the superimposed trends in EEG signals, which leads to poor performance of entropy algorithms in realistic EEG applications. Thus, we developed a multi-scale inherent fuzzy entropy algorithm [28] that has been applied in a previous resting-state migraine study [29]. In this SSVEP-based study, we modified the original version of the algorithm to a multi-scale relative inherent fuzzy entropy algorithm aiming to eliminate individual differences.

In addition to testing the Inherent FuzzyEn algorithm, we employed the same data processing steps using other entropy algorithms, one with a fuzzy structure (FuzzyEn) and one with a non-fuzzy structure (SampEn and ApEn). Considering healthy subjects who received repetitive visual stimulations, our findings showed that the entropy algorithms with fuzzy structures (Inherent FuzzyEn and FuzzEn) exhibited good performance compared with algorithms with non-fuzzy structures (SampEn and ApEn). Furthermore, the performance of the Inherent FuzzyEn algorithm was superior to the performances of the FuzzyEn, SampEn and ApEn models. In summary, the Inherent FuzzyEn algorithm was more effective in evaluating EEG signals from healthy subjects and migraine patients not only during resting-state but also during the repetitive SSVEP condition.

## VI. Conclusion

This study extracted SSVEP-based multi-scale relative inherent fuzzy entropy from migraine patients and HCs using a wearable headband EEG device. Our results highlight the feasibility of using a novel entropy measurement to compare EEG complexity in repetitive visual stimulation. Occipital EEG entropy showed an enhanced trend in patients in the inter-ictal phase; whereas, a reverse trend in patients in the pre-ictal phase. Additionally, patients in the pre-ictal phase exhibited a significantly lower transitional variation in EEG entropy than patients in the inter-ictal phase. Of note, we also noted that inherent fuzzy entropy was superior to other competing entropy models for conducting SSVEP experiments. In summary, inherent fuzzy entropy can be used in novel applications of visual stimulus environments for migraine studies, which may potentially be used in the future to provide a pre-ictal alert to migraine patients.

In this study, we have demonstrated the feasibility of the SSVEP-based complexity using a wearable headband EEG in support of detection of migraine attacks. The use of dry electrodes allows for easy and rapid monitoring on a daily basis and the advances in EEG recording and analysis ensure a promising future in support of individual solutions. We have ongoing advances in practical approaches of brain signal recording and sophisticated designs of extracting knowledge from neuro-information and home healthcare solutions are envisioned to guide to a wide range of real-life applications in the near future.